\documentclass[aps,prd,floats,nofootinbib,preprintnumbers,twocolumn]{revtex4}
\usepackage{graphicx}

\usepackage{amsmath}

\def\bea{\begin{eqnarray}}
\def\eea{\end{eqnarray}}

\newcommand{\be}{\begin{equation}}
\newcommand{\ee}{\end{equation}}
\newcommand{\mcl}[1]{\mathcal{#1}}

\begin{document}

\preprint{NUHEP-TH/14-01}
\preprint{UCI-HEP-TR-2013-20}

\title{Criteria for Natural Hierarchies}

\author{Andr\'e de Gouv\^ea}
\affiliation{Northwestern University, Department of Physics \& Astronomy, 2145 Sheridan Road, Evanston, IL~60208, USA}

\author{Daniel Hern\'{a}ndez}
\affiliation{Northwestern University, Department of Physics \& Astronomy, 2145 Sheridan Road, Evanston, IL~60208, USA}

\author{Tim M.P. Tait}
\affiliation{Department of Physics and Astronomy,
University of California, Irvine, CA 92697, USA}

\begin{abstract}
With the discovery of a particle that seems rather consistent with the minimal Standard Model Higgs boson, attention
turns to questions of naturalness, fine-tuning, and what they imply for physics beyond the Standard Model and its discovery 
prospects at run II of the LHC.  In this article we revisit the issue of naturalness, discussing some implicit 
assumptions that underly some of the most common statements, which tend to assign physical significance to 
certain regularization procedures. Vague arguments concerning fine-tuning can lead to conclusions that are 
too strong and perhaps not as generic as one would hope. Instead, we explore 
a more pragmatic definition of the hierarchy problem that 
does not rely on peeking beyond the murky boundaries of quantum field theory: we investigate the fine-tuning of the electroweak 
scale associated with thresholds from heavy particles, which is both calculable and dependent on the nature of the would-be 
ultraviolet completion of the Standard Model. 
We discuss different manifestations of new high-energy scales
that are favored by experimental hints for new physics 
with an eye 
toward making use of fine-tuning in order to determine natural regions of the new physics parameter spaces. 
\end{abstract}

\maketitle

\setcounter{equation}{0}
\setcounter{footnote}{0}

\section{Introduction}

The discovery of the Higgs boson at the Large Hadron Collider \cite{Aad:2012tfa,Chatrchyan:2012ufa} and the initial assessment that its properties are, at least approximately,  Standard Model (SM)-like \cite{Giardino:2012ww,Carmi:2012in,Plehn:2012iz}, renews interest on the question of whether electroweak symmetry breaking is natural \cite{Vissani:1997ys,Casas:2004gh,Casas:2006bd,Foot:2007ay,Foot:2007iy,Grinbaum:2009sk,Wetterich:2011aa,Bazzocchi:2012de,Farina:2013mla,Fabbrichesi:2013qca,Heikinheimo:2013fta,Dubovsky:2013ira,Giudice:2013yca,Feng:2013pwa,survey,Tavares:2013dga,Foot:2013hna}. Fundamental scalar fields are widely regarded as unnatural, a belief that has driven much of the theoretical exploration of TeV-scale physics beyond the Standard Model in the past few decades. 

At this point in time, it is appropriate to take stock of the different notions of naturalness, and to ask what naturalness {\em really} offers in the way of guidelines to new energy scales in physics, and what underlying assumptions were made along the way to inferring them.

In this manuscript, we critically explore the meaning of naturalness as embodied in the Higgs mass-squared parameter, and advocate a very concrete, unambiguous definition of fine-tuning. We use this definition to investigate generic extensions of the Standard Model. 





\section{The Hierarchy Problem}


The Standard Model implements spontaneous breaking of
$SU(2)_L \times U(1)_Y \rightarrow U(1)_{\rm EM}$ by positing the existence of the scalar Higgs field, an $SU(2)_L$ doublet with hyper-charge $+1/2$, whose potential is
\be
V \left( H \right)  =  - \mu^2 | H |^2 + \lambda | H |^4.
\ee
The Higgs potential is parameterized by a dimensionful mass-squared parameter $\mu^2$ and a dimensionless Higgs self-coupling $\lambda$. Together they set the Higgs vacuum expectation value (VEV) $v= \sqrt{\mu^2 / \lambda}=246$~GeV, which ultimately controls the masses of the $W$ and $Z$ bosons as well as that of the SM fermions (except, perhaps, the neutrinos). The parameters $\lambda$ and $v$ also set the mass of the physical Higgs boson (the Higgs particle), $M_h^2 = \lambda v^2$. Modulo a more complicated Higgs sector, the LHC measurement of the Higgs particle mass allows one to completely reconstruct the Higgs potential at the weak scale.

The hierarchy, fine-tuning, or naturalness problem refers to quantum corrections to the Higgs mass-squared parameter $\mu^2$ or, equivalently, the Higgs vacuum expectation value.\footnote{In 
this work we will always refer to quantum corrections to the mass-squared parameter $\mu^2$.}
A na\"{\i}ve description is as follows. Corrections to a scalar masses-squared are quadratically divergent and hence loops of Standard Model particles induce quantum corrections proportional to the unknown cutoff scale. The top quark, as the most strongly coupled SM particle to the Higgs field, will induce the most relevant such correction. Indeed, if one introduces a hard cutoff for the integral over the loop momentum, one finds, at one-loop,
\be
\delta \mu^2 = -\frac{3 \lambda_t^2}{8 \pi^2}~\times~\Lambda^2,
\label{eq:top}
\ee
where $\lambda_t = \sqrt{2}m_t / v$ is the top Yukawa coupling and $\Lambda$ is the cutoff scale.
In order to arrive at the physically observed value of $\mu^2$, one cancels this contribution
by adjusting the tree-level parameter such that the desired value is obtained.  Since there is no
theoretical structure suggesting a relationship between $\mu^2$, $\Lambda$, and $\lambda_t$, 
$\delta \mu^2 \gg \mu^2$ (or in other words, $\Lambda \gg $~TeV),
implies an unnatural fine-tuning of the counter-term.

This argument invests the regulator with physical meaning. 
For example, if space-time were in fact a lattice of points with
spacing $1/\Lambda$,
one would expect the mass-squared $m^2$ of a scalar field living on the lattice sites to receive finite corrections similar to 
 Eq.~(\ref{eq:top}),
and the theory would indeed appear to be finely tuned if that spacing were much smaller than $1 /  m$.
Of course, this interpretation -- and choice of regulator --  is hardly unique.
One could have alternately decided to regulate using dimensional regularization \cite{'tHooft:1972fi}, in which
case there would have been no quadratic sensitivity associated with the regulator, and the divergences would instead
manifest themselves as poles of Gamma functions as the dimensionality of space-time approaches four (or two).  One
could, very na\"{\i}vely, speculate that space-time is in fact fractal,\footnote{We choose not
to discuss this possibility in more detail, but it is worth mentioning that
experimental limits from orbital dynamics and the Lamb shift require
$\epsilon \lesssim 10^{-11}$ \cite{Schafer:1986oda}.} of dimensionality $4 + \epsilon$. In this case,
since the Standard Model contains only one single dimensionful parameter,
$\delta \mu^2$ is proportional to $\mu^2$ \cite{Bardeen:1995kv}.
Indeed, since $M_h$ seems as though it
may be consistent with vacuum stability \cite{EliasMiro:2011aa} and is below the triviality bound, the SM
could be well-behaved up to Planck scale energies (where gravitational effects become important), in which
case it is precisely the {\em absence} of heavy new states (coupled to the Higgs)
which allows the theory to be natural. We elaborate more on this point below.

A more standard interpretation is that the cutoff $\Lambda$ discussed above represents heavy states which have been integrated out of
the theory, leaving behind finite corrections to $\mu^2$ proportional to the square of the new heavy mass. 
This is most often what is implicitly {\em meant} by Eq.~(\ref{eq:top}).  More properly,
a heavy mass scale feeds into $\mu^2$ through a finite correction, and is independent of the choice of
regularization scheme.  The identification of the quadratic sensitivity of $\mu^2$ to the heavy scale with
a cutoff implicitly defines an effective theory (EFT) that is valid below the mass of the heavy state, and
the bare value of
$\mu^2$ in the EFT will differ from the value in the full theory by something of order the heavy mass-squared.
Of course, the details matter, and they depend not only on the new physics scale, but also on {\em whether and how} it ``talks'' to the Higgs boson.

It is known that the SM is not the complete theory of Nature.  For one thing,
it is missing descriptions of gravitation, dark matter, neutrino masses, and inflation.
In addition,
the fact that several of its couplings have Landau poles indicates that the physics must change at
very short distances for internal consistency.
In practice, however, Landau poles occur around or above the Planck scale, where the SM must be supplemented
by a theory of quantum gravity, and will be henceforth ignored.

Since neutrino masses, dark matter and other experimentally driven puzzles do not necessarily imply the existence of new, heavy states, the existence of gravity is often sold as the ultimate source for the hierarchy problem. 
In the absence of a compelling ultraviolet (UV) completion for gravity, however, it is impossible to rigorously compute, or even estimate, corrections
to $\mu^2$ and argue how they scale with the Planck scale, $M_{Pl}$. 
While there
is an effective field theory that properly describes gravitational interactions at energy scales well below the
Planck scale,
above $M_{Pl}$
there is no uniquely
established consistent quantum field theoretical formalism capable of describing the
short distance gravitational
interactions of the various SM fields.
It is widely believed that a realistic UV completion
will involve new states with Planck scale masses
(though see Refs.~\cite{Zichichi:1978gb,Shaposhnikov:2009pv} for counter-examples), and these
would seem likely to contribute to $\mu^2$ in analogy with Eq.~(\ref{eq:top}).

Even if one is willing, however, to accept that the quantum theory of gravity involves such states,
it is worthwhile to examine the assumptions behind the usual arguments.
For example, despite the fact that quantum gravitational effects {\em will} almost certainly
modify the short distance
interactions between the Higgs and the top quark, there is no generic reason to believe that 
corrections to the Higgs mass-squared parameter are in any way related to the strength of the top Yukawa interaction.  
The finite corrections to $\mu^2$ induced by integrating out such states involve {\em their} couplings to the Higgs,
and may have nothing to do with $\lambda_t$.\footnote{It is, however, worth noting
that specific UV-completions do contain states, like KK modes and excited string
resonances of the top quark, which are likely to have this property.}
As a result, there is no reason to think that the top is in any way more special than any of the other
SM fields in terms of its correction to $\mu^2$.
This is an important point, because one of the
usually claimed
hallmarks of a natural theory is the existence
of ``top partners." Instead, we are arguing that their presence is not at all generic!

In summary, it can be misleading to think that the quadratic divergences which plague the Higgs boson
mass-squared parameter in the SM are indicative that there is new physics at the TeV scale. The correct picture is, in some
sense, the opposite: {\em If} there is new physics beyond the SM {\em and} it ``talks" directly to the Higgs
sector, the new physics is likely to introduce large finite quantum corrections to the Higgs boson mass-squared. In order to
avoid a very unnatural fine-tuning, one is led to conclude that
the new physics scale cannot be larger than the TeV {\em if} it
couples to the Higgs boson through order one couplings. In the next few sections, we discuss different types of new physics and what the fine-tuning 
problem has to say about would-be new physics parameters, including the new mass scale.


\section{Heavy Fermion}
\label{fermion}

To begin with, we consider theories extending the SM degrees of freedom by at least one new heavy fermion,
generically denoted $\Psi$, which could be either
Majorana or Dirac (we consider both cases below, where relevant).
We are not concerned with the origin of the mass scale of this new state, $M_\Psi$, but instead concentrate on
the question:   under what circumstances is the SM + $\Psi$ model natural? 
The answer obviously depends on how $\Psi$ interacts with the SM,  and 
we consider several motivated possibilities in turn.


\subsection{Uncoupled $\Psi$}

The simplest scenario, arguably, is to assume that the Lagrangian ${\cal L}_{SM+\Psi}$ cleanly splits 
into ${\cal L}_{SM}+{\cal L}_{\Psi}$. 
In this case, there is no loop correction to the Higgs boson mass-squared parameter mediated 
by the interactions in ${\cal L}_{SM+\Psi}$ that involves $M_{\Psi}$. However, once we augment 
${\cal L}_{SM+\Psi}$ to include gravitational interactions, this is no longer the case.

\begin{figure}
\begin{center}
\hspace*{-0.75cm}
\includegraphics[width=0.22\textwidth]{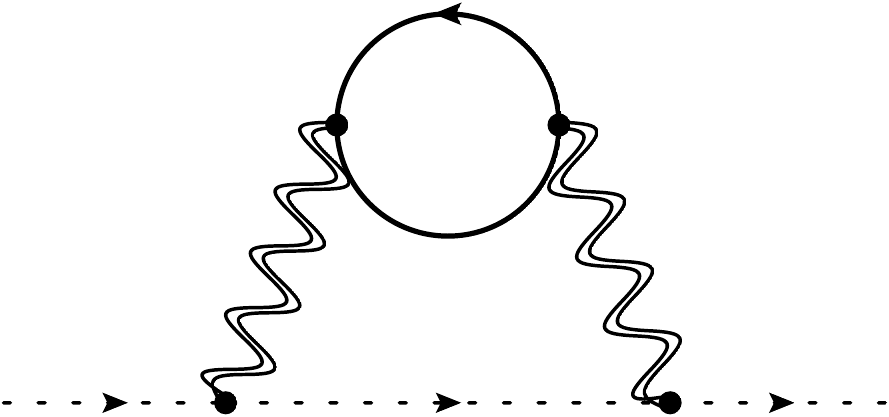} \hspace*{0.3cm}
\includegraphics[width=0.22\textwidth]{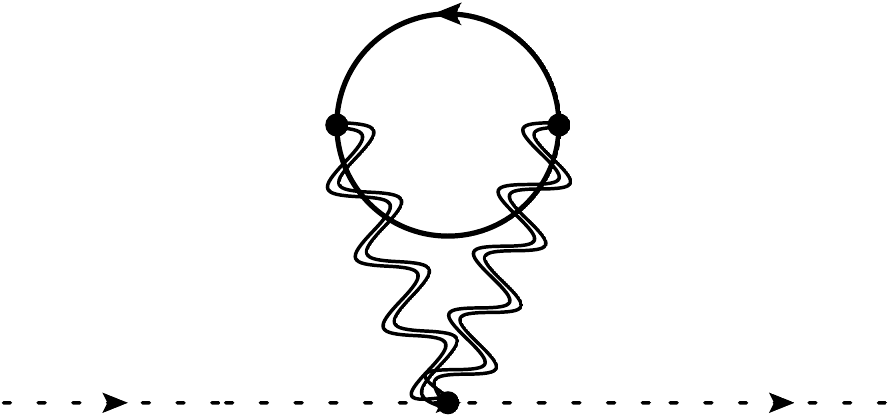}
\includegraphics[width=0.22\textwidth]{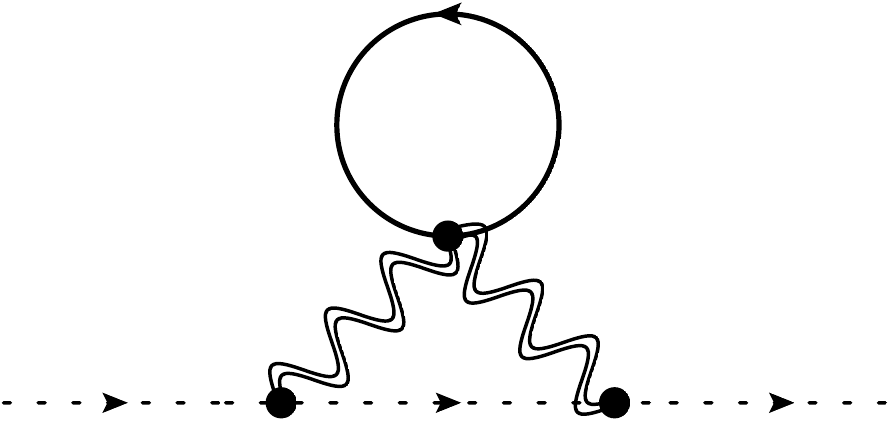} \hspace*{0.3cm}
\includegraphics[width=0.22\textwidth]{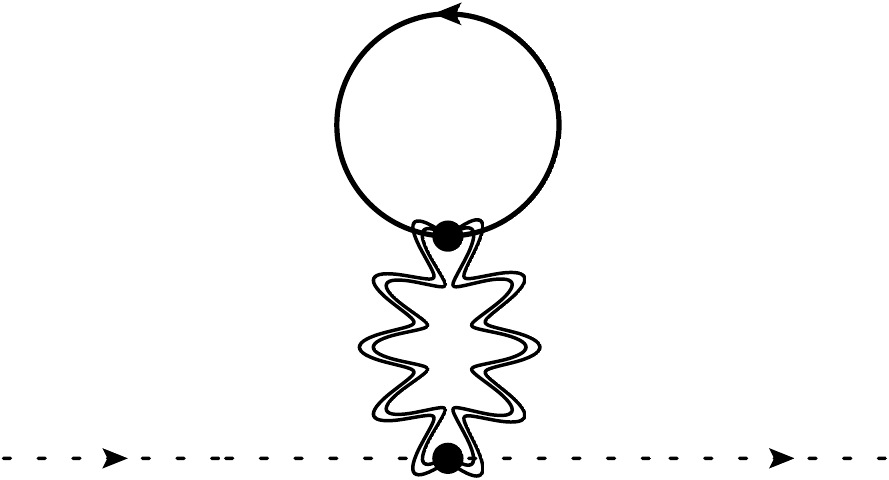}
\caption{Representative Feynman diagrams for 
corrections to the Higgs boson (dotted line) mass-squared  from a heavy 
particle (solid line) due to gauge boson or graviton exchange (double wavy line).}
\label{fig:gravity}
\end{center}
\end{figure}

While the full quantum theory of gravity is unknown, it is easy to examine how perturbative (low energy)
gravitational interactions mediate corrections to $\mu^2$ that are proportional to the new physics scale 
$M_{\Psi}$.  At the lowest order (two loops), diagrams, such as those shown in Figure~\ref{fig:gravity},
will induce finite corrections of order $1 / M_{Pl}^4$, the largest of which are proportional to
\be
\delta \mu^2  \sim  \frac{1}{\left( 16 \pi^2 \right)^2} \frac{M_{\Psi}^4}{M_{Pl}^4} ~\times~ \mu^2.
\label{twoloop}
\ee
These finite corrections are proportional to 
$\mu^2$ itself, and thus will not destabilize the weak scale even if $M_{\psi}$ is of order the Planck scale. 
The $\mu^2$ dependency is a consequence of the fact that the graviton coupling to a massless, on-shell particle with zero momentum vanishes. 

\begin{figure}
\begin{center}
\hspace*{-0.5cm}
\vspace*{-1.5cm}
\includegraphics[width=0.3\textwidth]{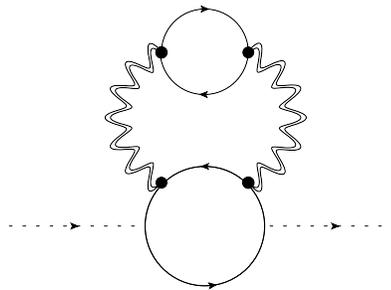}
\caption{Representative three-loop mixed gravity--SM Feynman diagram that leads to a  finite
correction to the Higgs boson (dotted line) mass-squared  from a heavy 
particle (solid line that couples only to gravity) due to graviton exchange (double wavy line) and top-quark exchange (solid line that couples to the Higgs boson).}
\label{fig:gravity3}
\end{center}
\end{figure}

Mixed gravity--SM loops, however, lead to potentially larger corrections, even for new heavy particles that coupled to all SM fields only through gravity.
Indeed, there are three-loop diagrams like the one depicted in Fig.~\ref{fig:gravity3}, where the Higgs divides into a pair of virtual top quarks, and the gravitons bridge from the top lines to a loop of the heavy fermion. In this case, all graviton vertices involve virtual particles, leading to a
stronger dependence on the heavy mass scale.  We estimate the correction from diagrams such as this to be,
\begin{equation}
\delta \mu^2  \sim  \frac{\lambda_t^2}{\left( 16 \pi^2 \right)^3} \frac{M_{\Psi}^4}{M_{Pl}^4} ~\times~ M_{\Psi}^2.
\label{threeloop}
\end{equation}
For $M_{\Psi}^2\gg\mu^2$, Eq.~(\ref{threeloop}) will significantly dwarf Eq.~(\ref{twoloop}). The naturalness constraint $\delta\mu^2\lesssim 100^2$~GeV$^2$ translates into $M_{\Psi}\lesssim 10^{14}$~GeV -- significantly smaller than the Planck scale but much larger than the weak scale.  

In summary: if ``quantum gravity'' effects can be 
parameterized by new heavy fermions with masses $\lesssim M_{Pl}$
that only couple gravitationally, our result seems to indicate that the SM 
plus ``quantum gravity'' is a moderately finely-tuned theory. New heavy fermions with masses below $10^{14}$~GeV can, however, co-exist peacefully with the weak scale if these only couple to the SM gravitationally.

\subsection{SM-Charged $\Psi$}

Next we consider the scenario where $\Psi$ is charged under $SU(2)_L\times U(1)_Y$. 
In this case, virtual $\Psi$ effects will (in the least) modify $\mu^2$ at the two-loop 
level,\footnote{Note that an electroweak singlet $\Psi$ carrying $SU(3)_C$ charge 
will still contribute at the three-loop level.} 
via the diagrams depicted in the top row of Fig.~\ref{fig:gravity}, where the graviton is 
reinterpreted as an $SU(2)_L\times U(1)_Y$ gauge boson. 
These diagrams contain a finite contribution to the Higgs mass-squared proportional to $M_\Psi^2$ \cite{Farina:2013mla}:
\be
\delta \mu^2 = \left(\frac{g^2}{16\pi^2}\right)^2  \times F\left(\frac{M_{W,Z}^2}{M_\psi^2}\right) \times M_\Psi^2,
\ee
where $F$ is a dimensionless function, of order one, and $g$ stands for a generic $SU(2)_L\times U(1)_Y$ 
gauge coupling. For $g^2/16\pi^2 \lesssim{\cal O}(10^{-2})$, the absence of fine-tuning requires 
$M_{\Psi}\lesssim 10$~TeV.  

New electroweakly-coupled fermions are among the many viable WIMP candidates for dark 
matter \cite{Cirelli:2005uq}. Assuming these make up all the dark matter and taking into account all experimental 
constraints, one arrives at $M_{\Psi}\sim 5$~TeV. The result above indicates that such ``Minimal Dark Matter'' 
models are natural.

\subsection{Yukawa-Coupled $\Psi$}

Finally, we consider the possibility that $\Psi$ couples to the Higgs field through a 
Yukawa interaction involving one of the SM fermions $\psi$, ${\cal L}_{SM+\Psi}\supset y_{\rm new}(\psi H)\Psi$. 
Here, virtual $\Psi$ effects modify $\mu^2$ at the one-loop level, 
\be
\delta \mu^2 \sim C~\frac{y_{\rm new}^2}{16\pi^2} \times M_{\Psi}^2.
\ee
where $C$ is the color factor appropriate for $\Psi$ and $\psi$.
Naturalness imposes the constraint $y_{\rm new}M_{\Psi}\lesssim 1$~TeV: either the new fermion mass is 
at the TeV scale, or the new fermion is weakly coupled to the Higgs boson.

One of the most exciting developments in particle physics of the previous decade is the experimental confirmation 
that neutrinos have mass, clear evidence of physics beyond the SM. 
The simplest implementation is a dimension five operator \cite{Weinberg:1979sa},
\begin{equation}
{\cal L}_{\nu\nu}=-\frac{\lambda_{ij}}{2M_{\nu}}L^iHL^jH + H.c.
\label{eq:numass}
\end{equation}
It is remarkable that the only observable consequence of ${\cal L}_{\nu\nu}$, for large enough 
$\lambda_{ij}/M_{\nu}$, is to provide the neutrinos (after electroweak symmetry) with very small Majorana masses
encoded in the mass matrix $m_{ij}=\lambda_{ij} v^2 / M_{\nu}$.
If ${\cal L}_{\nu\nu}$ is the low-energy manifestation of the new physics responsible for massive neutrinos, 
$M_{\nu}$ is to be interpreted, loosely speaking, as the largest possible energy scale above which the SM is no longer 
valid.\footnote{We imagine setting the largest $\lambda_{ij}\equiv1$ in such a way that the mass scale represented by $M_{\nu}$ is well-defined.}   

The details as to how this new scale feeds into $\mu^2$
depends on the UV-completion of the operator in Eq.~(\ref{eq:numass}).  
Given the loose requirement that, at low energies, the new physics Lagrangian yields the SM plus 
${\cal L}_{\nu\nu}$, it is impossible to single out the correct Lagrangian. Here, we will concentrate on the simplest 
interpretation of ${\cal L}_{\nu\nu}$, the Type-I seesaw mechanism \cite{seesaw},  described by 
\begin{equation}
{\cal L}_{SM}+\bar{N}_i\bar{\sigma}^{\mu}\partial_{\mu} N^i-\frac{M^{ij}_{R}}{2}N_iN_j-y_{ij}L^iN^jH+H.c.,
\end{equation}
where $i,j$ are family indices, $M_R$ is the right-handed neutrino mass-matrix, and 
$y$ is the neutrino Yukawa coupling matrix. Upon integrating out the right-handed neutrino fields, 
${\cal L}_{\nu\nu}$ is generated, with $\lambda/M_{\nu}\equiv yM_R^{-1}y^{\top}$.

This is a concrete example of the general case discussed above, with $\psi=L_i$ and $\Psi=N_i$ (in this case there are at least two of them).
At one loop, the finite correction to $\mu^2$ is \cite{Casas:2004gh},
\be
\delta \mu^2  =  - \frac{1}{4 \pi^2} \sum_{ij} | y_{ij} |^2 \times M_j^2~.
\ee
If one were to assume that all of the right-handed neutrinos have the same mass $M$, the bound will be dominated by the 
combination of the $y$'s responsible for the heaviest of the SM neutrinos, which could be as light as
the mass scale characterizing atmospheric mixing, 0.05 eV, or as heavy as one third of the cosmological upper limit on the sum of the
neutrino masses, 0.7 eV \cite{deGouvea:2013onf}.  Under these simplifying assumptions, 
demanding that $\delta \mu^2 \lesssim v^2$ translates into the requirement,
\be
M \lesssim  8 - 14 \times 10^3~{\rm TeV}.
\ee
This is an interesting result, discussed earlier in \cite{Casas:2004gh,Farina:2013mla}, with non-trivial consequences. If the Type-I seesaw mechanism is realized in  Nature, the seesaw scale must be below $10^4$~TeV or so. On the one hand, because there always remains the possibility that the neutrinos are weakly coupled, the new physics scale need not be below a few TeV -- it can be some orders of magnitude higher. On the other hand, ``electroweak--natural'' seesaw scales are much lower than those required by almost all versions of leptogenesis. 




As another example, 
theories which attempt to explain the hierarchal structure apparent in the quark masses and mixing angles also often
invoke heavy particles.  As an example, we consider an implementation of the Froggatt-Nielsen mechanism
\cite{Froggatt:1978nt} which invokes a global $U(1)$
symmetry to forbid the Yukawa interactions of the light quarks.
This symmetry is broken by the VEV of a scalar particle carrying $U(1)$ charge, which induces mixing between
the SM fermions and a set of additional fermions which are vector-like under the electroweak interaction,
allowing for gauge-invariant masses.  For the purposes of our discussion, we limit ourselves to a subset of the model
describing the top and bottom quark masses,
\bea
{\cal L}\supset i \overline{D}  \hspace*{-1.5mm} \not \hspace*{-0.3mm} \partial D - M_D \overline{D} D
+ y_1 \overline{Q}_3 D H
+ y_2 \overline{D} b_R \phi + H.c.,
\eea
where $Q_3$ is the third generation quark doublet, $t_R$ and $b_R$ are the quark singlets, $\phi$ is the scalar
whose VEV breaks the $U(1)$ global symmetry, and $D$ is a vector-like quark with the same SM gauge interactions
as $b_R$.  If the $U(1)$ charges are chosen as: $Q_3: +1$, $t_R: +1$, $D: +1$, $b_R: 0$, $H: 0$, $\phi: +1$, the top
Yukawa will be directly allowed by the symmetry whereas the bottom 
Yukawa is induced only after integrating out $D$,
\bea
y_b^{\rm eff} & = & y_1 y_2 \frac{\langle \phi \rangle}{M_D}~.
\eea
Expanding this picture to include additional vector-like fermions (all with similar masses)
with ${\cal O}(1)$ $U(1)$ charges allows one to reproduce the entire
structure of the observed quark masses and mixings, invoking only ${\cal O}(1)$ couplings $y_i$
\cite{Leurer:1992wg}.

The existence of the $y_1$ interaction implies that the $D$ mass will contribute to the Higgs potential.
There will be mixed loops involving the $D$ fermion and the third generation doublet which are structurally
very similar to the correction from the neutrino singlet discussed above.  The resulting correction,
\bea
\delta \mu^2 & = & - \frac{6 |y_1|^2}{8 \pi^2} \times M_D^2,
\label{eq:Hflav}
\eea
will result in $\delta \mu^2 \gtrsim v^2$ unless,
\bea
M_D & \lesssim & \frac{\rm 900~GeV}{|y_1|}~.
\eea
Quantum corrections to $\mu^2$ proportional to 
$M_{\phi}^2$ are also generically expected. We discuss these in detail in the next section.  

The exchange of the heavy $D$ quarks and $\phi$ also induce flavor-violating effects. These are in conflict with observations unless the masses of the new states are $\gtrsim 1000$~TeV \cite{Isidori:2010kg}.  In the absence of additional
ingredients, the flavor-violating constraints together with naturalness (i.e., demanding $\delta \mu^2 \lesssim v^2$) 
imply $y_1 \lesssim 10^{-3}$. Naturalness combined with the absence of flavor-violating effects does 
not allow for a flavor model where all couplings are order one. 

In summary, if there are new fermions that are Yukawa-coupled to the Higgs boson, these are required to be light 
(with masses below $\sim 1$ TeV), or the new Yukawa couplings must be small,
or additional ingredients must be invoked to insure naturalness. 
If the Type-I seesaw is realized, right-handed neutrino masses are constrained to be less than 
several thousand TeV and, in the absence of more ``new physics,'' 
most versions of leptogenesis are ruled out by naturalness. 
Naturalness also provides nontrivial constraints for new physics interpretations of the flavor puzzles.

\section{Heavy Boson}
\label{scalar}

In parallel to the case of additional heavy fermions discussed in Section~\ref{fermion}, in this section we discuss
theories containing one or more heavy bosons in addition to the SM degrees of freedom.  Most of this discussion
will involve a heavy scalar (real or complex)
$\Phi$ of mass $M_\Phi$, but it also applies to massive gauge bosons whose masses
are the consequence of spontaneous symmetry breaking.  The primary question remains: under which
circumstances is the SM + $\Phi$ model natural?  As before, the answer depends on how $\Phi$ interacts. 

\subsection{$|H|^2 |\Phi|^2$ Coupling}

In contrast to
the fermion case and independently of the quantum numbers of $\Phi$, 
at least one marginal operator linking $\Phi$ to the Higgs and
consistent with all of the SM gauge symmetries is always allowed: 
\be
{\cal L}_{SM +\Phi}\supset \lambda_{\rm new} |H|^2|\Phi|^2. 
\label{eq:h2phi2}
\ee
At the one-loop level, this interaction allows the 
mass scale $M_\Phi$ to contribute to the Higgs boson mass-squared,
\be
\delta \mu^2  \sim  \frac{\lambda_{\rm new}}{16 \pi^2} \times M_{\Phi}^2. 
\ee
Similar to the Yukawa-coupled-fermion case, naturalness dictates that either the interaction strength 
is very small, or that $\Phi$ is very light, i.e., $M_{\Phi} \lesssim 1$~TeV. 

Several well-motivated models for new physics fall into this category. For example, in ``Higgs Portal"
dark matter \cite{Burgess:2000yq}, where the SM is augmented to include one gauge-singlet scalar 
that couples to the SM via Eq.~(\ref{eq:h2phi2}), the requirement that $\Phi$
is a thermal relic 
making up all of the dark matter places bounds on combinations of $\lambda_{\rm new}$ and $M_{\Phi}$. 
These translate into $M_{\Phi} < 1$~TeV, which means such models safely meet our naturalness criterium. 

Another  very well known example is to postulate that the SM is the low-energy remnant of a larger grand unified gauge theory (GUT), spontaneously broken at $M_{\rm GUT} \gg v$.  
The Higgs doublet will necessarily make up part of a larger GUT multiplet, and $\mu^2$ will receive corrections
from loops of the GUT bosons 
\bea
\delta \mu^2 & = & \frac{C}{16 \pi^2} ~\times~ M_{\rm GUT}^2, 
\eea
where $C$ is a coefficient of order (at least) the known gauge couplings that depends on the detailed physics at the GUT scale.  Since limits from
proton decay require $M_{\rm GUT} \gtrsim 10^{16}~{\rm GeV}$, it is clear that this correction to $\mu^2$ is highly
unnatural, and requires a
seemingly magical cancellation between the tree-level Higgs mass-squared parameter and all of its higher order quantum 
corrections\footnote{This fine-tuning arising from one-loop corrections to $\mu^2$ from GUT scale physics is
distinct from the (tree-level) doublet-triplet splitting problem.}.




Another very well motivated extension of the SM addresses the apparent smallness of the QCD $\theta$ angle
by promoting it to a dynamical field  \cite{Peccei:1977hh}.  
We illustrate the discussion by considering a simple model framework \cite{Kim:1979if,Shifman:1979if}
which extends the SM by a complex SM-singlet scalar field $\Phi$ which couples to a colored fermion $\psi$
\be
| \partial_\mu \Phi |^2 + M^2 |\Phi |^2 - \lambda_\rho |\Phi |^4 + i  \overline{\psi} \hspace*{-1.5mm} \not \hspace*{-0.3mm}
\partial \psi
+ i y \overline{\psi} \Phi \gamma_5 \psi + H.c.
\ee
This Lagrangian has a $U(1)_{PQ}$ symmetry, which is anomalous with respect
to the $SU(3)_c$ gauge symmetry.  Assuming $U(1)_{PQ}$ is  broken via the VEV 
$\langle \Phi \rangle = \sqrt{M^2 / \lambda_\rho} \equiv f_a$,
the phase of $\Phi$
emerges as the axion field, a pseudo-Goldstone boson whose expectation value cancels any
pre-existing coefficient of $G \widetilde{G}$, solving the strong CP problem.  The modulus of $\Phi$,
$\rho$, is a real scalar field whose mass is related to the $\Phi$ potential 
by $M_\rho^2 = \lambda_\rho \langle \Phi \rangle^2 = \lambda_\rho f_a^2$.
Astrophysical constraints from red giant stars and supernovae generically require $f_a \gtrsim 10^9$~GeV
\cite{Raffelt:1990yz}, guaranteeing that a realistic axion theory will involve heavy mass scales,
possibly destabilizing the Higgs potential.

The dangerous interaction is the mixed quartic involving the Higgs and $\Phi$, Eq.~(\ref{eq:h2phi2}).
This quartic interaction will lead, as already discussed, to a finite correction to the Higgs mass-squared parameter of the form,
\be
\delta \mu^2 \sim \frac{\lambda_{\rm new}}{16 \pi^2} \times M_\rho^2 = 
 \frac{\lambda_{\rm new} \lambda_\rho}{16 \pi^2} \times f_a^2~.
\ee
For $f_a \gg v$, this correction leads to electroweak fine-tuning, unless $\lambda_{\rm new}$ is very small.

Strictly speaking, the solution to the strong CP problem does not depend on any particular value of $\lambda_{\rm new}$, and it
would be tempting to simply discard it.  However, it can be forbidden by no symmetry, and receives additive corrections
from higher order processes.  As discussed in Ref.~\cite{Farina:2013mla}, there are contributions at two loops if the $\psi$
fermions are electroweakly charged, and at three loops if not.  The authors
of \cite{Farina:2013mla} interpret the size of such corrections
as minimal values of $\lambda_{\rm new}$ which thus characterize a minimal level of fine-tuning.

\setcounter{footnote}{0}

An alternate strategy would be to take $\lambda_\rho \ll 1$, reducing the mass of the $\rho$ 
scalar\footnote{$\lambda_{\rho}$ is also additively
corrected at one loop by the $\Phi$-$\overline{\psi}$-$\psi$ coupling.}.   Even for $\lambda_{\rm new} \simeq 1$, 
$\delta \mu^2 \lesssim v^2$ provided $M_\rho \lesssim 2$~TeV.  It is intriguing that such a particle 
(with order one couplings to the Higgs) could conceivably be within the grasp of future high energy colliders. 

\subsection{$|H|^2\Phi$ Coupling}

A gauge-singlet scalar can also couple singly to a pair of Higgs fields via
\be
{\cal L}_{SM +\Phi}\supset \kappa_{\rm new} |H|^2\Phi,
\label{eq:h2phi}
\ee
where $\kappa_{\rm new}$ is a coupling constant with dimensions of mass.  In the limit
$\kappa_{\rm new} \rightarrow 0$, there is an enhanced $Z_2$ symmetry under which
$\Phi$ is odd and the SM is even, indicating that any value of $\kappa_{\rm new}$ is
natural in the sense of 't Hooft.  A stable perturbative loop expansion requires
$\kappa_{\rm new} \lesssim M_h, M_\Phi$.

This coupling induces a correction to the Higgs mass-squared (for $M_\Phi \gg M_h$),
\be
\delta \mu^2  \sim  -\frac{\kappa_{\rm new}^2}{16 \pi^2} \times \log\left(\frac{M^2_\Phi}{M^2_h}\right) ,
\label{eq:dmuh2phi}
\ee
whose size is characterized by $\kappa_{\rm new}$, and depends only logarithmically on $M_\Phi$.
Obviously the theory will be finely tuned unless $\kappa_{\rm new} \lesssim$~TeV.  What is novel is the fact that
the super-renormalizable interaction effectively shields the Higgs mass from the heavy mass scale $M_\Phi$,
despite allowing for relatively large coupling between the Higgs and $\Phi$ sectors.

As in the previous subsection, the equivalent of the four-point coupling  $\lambda_{\rm new}$ of Eq.~(\ref{eq:h2phi2}) is expected to exist as it 
is consistent with all symmetries. It can, however, be chosen arbitrarily small, and no large corrections are expected. 
At one loop, the $\kappa_{\rm new}$ interaction results in a four-point interaction reminiscent of that of
$\lambda_{\rm new}$, of ${\cal O}(\kappa_{\rm new}^2/v^2)$.
However, this contribution does not require a counter-term, and the theory is self-consistent
even without including $\lambda_{\rm new}$ as a fundamental interaction. In a nutshell, the counter-terms to $\lambda_{\rm new}$
are proportional to $\lambda_{\rm new}$ itself. The upshot is that,
provided $\lambda_{\rm new} \ll 1$ and $\kappa_{\rm new} \lesssim$~TeV, the Higgs mass-squared parameter
receives no large corrections, even when $M_\Phi \gg v$.


This radiatively stable separation of the Higgs from heavy states can persist even for more complicated
choices of dark sector.  If one extends the construct by including a gauge-singlet fermion $\Psi$ whose
coupling to the SM is entirely through a Yukawa interaction, $y \Phi \bar{\Psi} \Psi$, where
$y$ is a dimensionless coupling.  This theory has a two loop contribution to $\delta \mu^2$ proportional
to $M_\Psi$,
\be
\delta \mu^2 = \frac{y^2 \kappa_{\rm new}^2 M_\Psi^2}
{(16\pi^2)^2 M_\Phi^2} ~ F\left( \frac{M_\Psi^2}{M_\Phi^2} \right)
\ee
where $F(M_\Psi^2/M_\Phi^2)$ is $\mcl{O}(1)$. The finite corrections proportional to $M_{\Psi}^2$ 
do not destabilize the weak scale as long as $M_\Psi^2 \lesssim M_\Phi^2$
(provided $\kappa_{\rm new} \lesssim$~TeV and $\lambda_{\rm new} \ll 1$).  
This feature
extends to diagrams with any number of loops, since in the limits $\kappa_{\rm new} \rightarrow 0$ or 
$M_\Phi \rightarrow \infty$, the SM and $\Psi$ must decouple. Thus, all contributions to $\delta \mu^2$ 
involving that sector should be equal to $\kappa_{\rm new}^2$ 
multiplied by an analytic function  of the ratio $M_\Psi^2 / M_\Phi^2$.

This type of construction suggests a natural model of heavy fermionic dark matter (played by $\Psi$ in the discussion
above) which communicates primarily with the Higgs via exchange of $\Phi$.   At low energies, $\Phi$ exchange
results in an operator of the form $|H|^2\bar{\Psi}\Psi$.  Such a coupling looks dangerous from the point of view of 
naturalness, but we have seen that a reasonable UV completion exists that is relatively free from fine-tuning.




\section{Summary and Conclusions}

Now that the Higgs has been discovered and appears at least roughly consistent with its expected SM properties, 
it is appropriate to consider what it means for the electroweak symmetry-breaking scale to be ``natural".  
Commonly made arguments based on loops of top quarks assign a physical meaning to the cutoff employed as a 
regulator in certain regularization procedures.  This could turn out to be a useful guide if, 
for example, space-time resembles a 
lattice at short distances or quantum gravity involves stringy resonances of the top.  On the other hand,
they could also very well turn out to be misleading.
If such assumptions are abandoned, the presence of top partner fields as
harbingers of naturalness, for example, loses much of its motivation.

A less ambiguous but very concrete measure of fine-tuning in the Higgs potential makes reference to the finite corrections
induced by integrating out heavy particles.  Such corrections are independent of the choice of regulator, 
and make no assumptions about physics beyond quantum field theory. 
Indeed, the question we largely address is the following: Under which circumstances can the SM augmented by
heavy fields accommodate, without large fortuitous cancellations, multiple mass scales? Our main message is that 
the answer depends on the nature of the new heavy particles and how they ``talk'' to the Higgs field.

The simplest solution, arguably, would be to assume that the weak scale is the only scale of high energy physics. 
In the absence of new, heavy particles, the SM, as is, is natural. Fortunately, 
Nature has already revealed that there is physics beyond the standard model. 
Dark matter and nonzero neutrino masses require new degrees of freedom and, perhaps, 
new mass scales. More indirect hints like the unification of gauge couplings and the fermion particle 
content (GUTs), the need for a mechanism of baryogenesis, the strong CP problem, and the flavor puzzle 
also suggest the existence of new, usually very heavy, new states.
Ultimately, the need to describe gravitation provides a concrete ultra-large mass scale, 
though its connection to massive particles is less clear.  

Grand unified theories, by themselves, certainly signal extreme fine-tuning.  
On the opposite end of the ``new physics'' spectrum, particles coupled to the 
SM only via gravity induce perturbative corrections which 
are relatively benign, even if the new particle masses are relatively close to the 
Planck scale ($M_{\rm new}\lesssim 10^{14}$~GeV).
Potentially weakly coupled extensions of the SM, such as the seesaw theory of neutrino masses, 
do not necessarily destabilize the weak scale, as long as the new physics scale is not arbitrarily high.  
In the case of the Type-I seesaw, the theory becomes unnatural if the right-handed neutrino masses are much
larger than about 1000 TeV, a perfectly viable alternative even if it spells doom for most manifestations of 
leptogenesis. The existence of a fundamental particle as thermal dark matter is natural as long 
as the dark matter mass is below tens of TeV, as is already suggested by the
unitarity bound.

``More'' new physics allows for the coexistence of dramatically different mass scales. 
A very popular solution is to postulate that the theory is supersymmetric at high energy scales (e.g., the GUT scale), 
in which case there will be additional contributions to $\mu^2$ from the super-partners, whose couplings are 
guaranteed by supersymmetry (SUSY) to lead to a combined $\delta \mu^2=0$.  Of course, 
SUSY cannot be an exact symmetry of Nature -- it must, somehow, be broken. As is well known, if SUSY 
is only softly-broken several of its desirable non-renormalization properties are maintained, 
shielding $\mu^2$ from the $M_{\rm GUT}^2$ corrections. 

The existence of the super-partners induces its own hierarchy problem. 
Soft SUSY breaking parameters (generically denoted as $\tilde{m}^2$) 
lead to finite quantum corrections to the Higgs boson mass-squared
\begin{equation}
\delta \mu^2 \sim\frac{\lambda^2}{16 \pi^2} ~ \times ~ \tilde{m}^2,
\end{equation}
meaning that if Nature is indeed (softly-broken) supersymmetric, some of the soft-breaking parameters 
(for example, the stop masses) must be around 1~TeV, in order to avoid incomprehensibly large quantum corrections 
to the Higgs mass-squared.  Broken supersymmetry creates the hierarchy problem, and then solves it 
for weak scale breaking parameters.

The fact that the gauge couplings coincide at $M_{\rm GUT}\simeq 10^{16}$~GeV if one assumes a 
supersymmetric version of the SM softly-broken at the TeV scale, together with the peculiar 
hyper-charge assignments of the matter fields makes the SUSY GUT picture quite attractive, and certainly does 
motivate SUSY at the electroweak scale.
In this case, the chain of reasoning regarding the hierarchy problem is that the GUT scale 
induces enormous corrections to the Higgs mass-squared parameter, which are tamed by SUSY, inducing new 
contributions from the SM super-partners (in particular the stop), 
which themselves demand that the stop masses are at the TeV 
scale. If the indirect evidence for a GUT turns out to be a red herring, the arguments in favor of 
SUSY become much more ambiguous.

Naturalness is a powerful motivating force behind the search for new 
TeV-scale degrees of freedom.  Understanding the assumptions
that underly its application and its precise formulation both serves as a guide to possible futures as well as
a primer for decoding (hopefully!) soon-to-be-made discoveries.

\section*{Acknowledgements}

TMPT is pleased to acknowledge discussions with Tom DeGrand, Jonathan Feng, Arvind Rajaraman, and
Flip Tanedo.  AdG and TMPT are grateful to the theory group of Fermilab, which nurtured their careers
when discussion concerning this work first began, and acknowledge many discussions with Bill Bardeen regarding the naturalness problem.
We thank Witold Skiba for discussions of our gravity loop estimate, and are 
especially indebted to Cliff Cheung for bringing to our attention the importance of the 
three-loop gravity diagram, which was neglected in an earlier version of this manuscript.
The research of TMPT is supported in part by NSF grant PHY-1316792
and by the University of California, Irvine through a Chancellor's Fellowship. The work of AdG and DH is sponsored in part by the DOE grant \#DE-FG02-91ER40684.


\begin{thebibliography}{99}
 
\bibitem{Aad:2012tfa} 
  G.~Aad {\it et al.}  [ATLAS Collaboration],
  Phys.\ Lett.\ B {\bf 716}, 1 (2012)
  [arXiv:1207.7214 [hep-ex]].
  
\bibitem{Chatrchyan:2012ufa} 
  S.~Chatrchyan {\it et al.}  [CMS Collaboration],
  Phys.\ Lett.\ B {\bf 716}, 30 (2012)
  [arXiv:1207.7235 [hep-ex]].
  
\bibitem{Giardino:2012ww} 
  P.~P.~Giardino, K.~Kannike, M.~Raidal and A.~Strumia,
  JHEP {\bf 1206}, 117 (2012)
  [arXiv:1203.4254 [hep-ph]].
  
\bibitem{Carmi:2012in} 
  D.~Carmi, A.~Falkowski, E.~Kuflik, T.~Volansky and J.~Zupan,
  JHEP {\bf 1210}, 196 (2012)
  [arXiv:1207.1718 [hep-ph]].
  
\bibitem{Plehn:2012iz} 
  T.~Plehn and M.~Rauch,
  Europhys.\ Lett.\  {\bf 100}, 11002 (2012)
  [arXiv:1207.6108 [hep-ph]].
  

\bibitem{Vissani:1997ys} 
  F.~Vissani,
  Phys.\ Rev.\ D {\bf 57}, 7027 (1998)
  [hep-ph/9709409].
  
\bibitem{Casas:2004gh} 
  J.~A.~Casas, J.~R.~Espinosa and I.~Hidalgo,
  JHEP {\bf 0411}, 057 (2004)
  [hep-ph/0410298].
  
\bibitem{Casas:2006bd} 
  J.~A.~Casas, J.~R.~Espinosa and I.~Hidalgo,
  Nucl.\ Phys.\ B {\bf 777}, 226 (2007)
  [hep-ph/0607279].
  
\bibitem{Foot:2007ay} 
  R.~Foot, A.~Kobakhidze, K.~.L.~McDonald and R.~.R.~Volkas,
  Phys.\ Rev.\ D {\bf 76}, 075014 (2007)
  [arXiv:0706.1829 [hep-ph]].
  
\bibitem{Foot:2007iy} 
  R.~Foot, A.~Kobakhidze, K.~L.~McDonald and R.~R.~Volkas,
  Phys.\ Rev.\ D {\bf 77}, 035006 (2008)
  [arXiv:0709.2750 [hep-ph]].
  
\bibitem{Grinbaum:2009sk} 
  A.~Grinbaum,
  Found.\ Phys.\  {\bf 42}, 615 (2012)
  [arXiv:0903.4055 [physics.hist-ph]].
  
\bibitem{Wetterich:2011aa} 
  C.~Wetterich,
  Phys.\ Lett.\ B {\bf 718}, 573 (2012)
  [arXiv:1112.2910 [hep-ph]].
  
\bibitem{Bazzocchi:2012de} 
  F.~Bazzocchi and M.~Fabbrichesi,
  Phys.\ Rev.\ D {\bf 87}, no. 3, 036001 (2013)
  [arXiv:1212.5065 [hep-ph]].

\bibitem{Farina:2013mla} 
  M.~Farina, D.~Pappadopulo and A.~Strumia,
  JHEP {\bf 1308}, 022 (2013)
  [arXiv:1303.7244 [hep-ph]].
  
\bibitem{Fabbrichesi:2013qca} 
  M.~Fabbrichesi and S.~Petcov,
  arXiv:1304.4001 [hep-ph].
  
\bibitem{Heikinheimo:2013fta} 
  M.~Heikinheimo, A.~Racioppi, M.~Raidal, C.~Spethmann and K.~Tuominen,
  arXiv:1304.7006 [hep-ph].
  
\bibitem{Dubovsky:2013ira} 
  S.~Dubovsky, V.~Gorbenko and M.~Mirbabayi,
  JHEP {\bf 1309}, 045 (2013)
  [arXiv:1305.6939 [hep-th]].
  
\bibitem{Giudice:2013yca} 
  G.~F.~Giudice,
  arXiv:1307.7879 [hep-ph].
  
\bibitem{Feng:2013pwa} 
  J.~L.~Feng,
  arXiv:1302.6587 [hep-ph].

  \bibitem{survey}
  {\tt http://workshops.ift.uam-csic.es/WMH126/survey.html}.
  
\bibitem{Tavares:2013dga} 
  G.~Marques Tavares, M.~Schmaltz and W.~Skiba,
  arXiv:1308.0025 [hep-ph].
  
\bibitem{Foot:2013hna} 
  R.~Foot, A.~Kobakhidze, K.~L.~McDonald and R.~R.~Volkas,
  arXiv:1310.0223 [hep-ph].

\bibitem{'tHooft:1972fi} 
  G.~'t Hooft and M.~J.~G.~Veltman,
  Nucl.\ Phys.\ B {\bf 44}, 189 (1972).

\bibitem{Schafer:1986oda} 
  A.~Schafer and B.~Muller,
  J.\ Phys.\ A {\bf 19}, no. 18, 3891 (1986).

\bibitem{Bardeen:1995kv} 
  W.~A.~Bardeen,
  FERMILAB-CONF-95-391-T.
  
  
\bibitem{EliasMiro:2011aa} 
  J.~Elias-Miro, J.~R.~Espinosa, G.~F.~Giudice, G.~Isidori, A.~Riotto and A.~Strumia,
  Phys.\ Lett.\ B {\bf 709}, 222 (2012)
  [arXiv:1112.3022 [hep-ph]].
 
\bibitem{Zichichi:1978gb} 
S.~Weinberg, contribution to:
  A.~Zichichi,
  ``Understanding the Fundamental Constituents of Matter. Proceedings: 1976 International School of Subnuclear Physics (NATO-MPI-MRST Advanced Study Institute), Erice, Trapani, Sicily, Jul 23-August 8, 1976,''
  New York, Usa: Plenum Pr.(1978) 915 P.(The Subnuclear Series, 14)
  
\bibitem{Shaposhnikov:2009pv} 
  M.~Shaposhnikov and C.~Wetterich,
  Phys.\ Lett.\ B {\bf 683}, 196 (2010)
  [arXiv:0912.0208 [hep-th]].
   
\bibitem{Cirelli:2005uq} 
  M.~Cirelli, N.~Fornengo and A.~Strumia,
  Nucl.\ Phys.\ B {\bf 753}, 178 (2006)
  [hep-ph/0512090].
  
  
\bibitem{Weinberg:1979sa} 
  S.~Weinberg,
  Phys.\ Rev.\ Lett.\  {\bf 43}, 1566 (1979).
  
  
   \bibitem{seesaw} P.~Minkowski, Phys.\ Lett.\ B {\bf 67}, 421 (1977);
M. Gell-Mann, P. Ramond and R. Slansky in {\it Supergravity},  eds. D. Freedman and P. Van Niuwenhuizen (North Holland, Amsterdam, 1979), p.~315;
T. Yanagida in {\it Proceedings of the Workshop on Unified Theory and Baryon Number in the Universe}, eds. O.~Sawada and A.~Sugamoto (KEK, Tsukuba, Japan, 1979); 
S.L.~Glashow, {\it 1979 Carg\`ese Lectures in Physics -- Quarks and Leptons}, eds. M.~L\'evy {\it et al.} (Plenum, New York, 1980), p.~707; 
R.N. Mohapatra and G. Senjanovi\'c, Phys.\ Rev.\ Lett.\ {\bf 44}, 912 (1980);
J.~Schechter and J.W.F.~Valle,  Phys.\ Rev.\  D {\bf 22}, 2227 (1980).

  
\bibitem{deGouvea:2013onf} 
  A.~de Gouv\^ea {\it et al.}  [Intensity Frontier Neutrino Working Group Collaboration],
  arXiv:1310.4340 [hep-ex].
  
  
\bibitem{Froggatt:1978nt} 
  C.~D.~Froggatt and H.~B.~Nielsen,
  Nucl.\ Phys.\ B {\bf 147}, 277 (1979).
  
\bibitem{Leurer:1992wg} 
  M.~Leurer, Y.~Nir and N.~Seiberg,
  Nucl.\ Phys.\ B {\bf 398}, 319 (1993)
  [hep-ph/9212278].
  
\bibitem{Isidori:2010kg} 
  G.~Isidori, Y.~Nir and G.~Perez,
  Ann.\ Rev.\ Nucl.\ Part.\ Sci.\  {\bf 60}, 355 (2010)
  [arXiv:1002.0900 [hep-ph]].
 
  
\bibitem{Burgess:2000yq} 
  C.~P.~Burgess, M.~Pospelov and T.~ter Veldhuis,
  Nucl.\ Phys.\ B {\bf 619}, 709 (2001)
  [hep-ph/0011335].
   
   

\bibitem{Peccei:1977hh} 
  R.~D.~Peccei and H.~R.~Quinn,
  Phys.\ Rev.\ Lett.\  {\bf 38}, 1440 (1977).
  
\bibitem{Kim:1979if} 
  J.~E.~Kim,
  Phys.\ Rev.\ Lett.\  {\bf 43}, 103 (1979).
  
\bibitem{Shifman:1979if} 
  M.~A.~Shifman, A.~I.~Vainshtein and V.~I.~Zakharov,
  Nucl.\ Phys.\ B {\bf 166}, 493 (1980).
  
\bibitem{Raffelt:1990yz} 
  G.~G.~Raffelt,
  Phys.\ Rept.\  {\bf 198}, 1 (1990).
  
   
 
 
 \end{thebibliography}
 \end{document}